\documentclass{llncs}
\usepackage{amssymb}
\usepackage{epsfig}
\usepackage{makeidx}  

\begin{document}

\begin{frontmatter}

\pagestyle{headings}  

\mainmatter

\title{Statistical User Model for the Internet Access}

\titlerunning{Statistical User Model for the Internet Access}  

\author{Carmen Pellicer-Lostao\inst{1} \and Daniel Morato\inst{2} \and Ricardo L\'{o}pez-Ruiz\inst{3}}

\authorrunning{L\'{o}pez-Ruiz and Morato and Pellicer-Lostao }   

\institute{Department of Computer Science, \\
Universidad de Zaragoza, 50009 - Zaragoza, Spain,\\
\email{carmen.pellicer@unizar.es}
\and
Department of Automatics and Computer Science, \\
Universidad P\'{u}blica de Navarra, 31006 - Pamplona, Spain,\\
\email{daniel.morato@unavarra.es}
\and
Department of Computer Science and BIFI, \\
Universidad de Zaragoza, 50009 - Zaragoza, Spain.\\
\email{rilopez@unizar.es} }

\maketitle              

\begin{abstract}
A new statistical based model approach to characterize a user's
behavior in an Internet access link is presented. The real
patterns of Internet traffic in a heterogeneous Campus Network are
studied. We find three clearly different patterns of individual
user's behavior, study their common features and group particular
users behaving alike in three clusters. This allows us to build a
probabilistic mixture model, that can explain the expected global
behavior for the three different types of users. 
We discuss the implications of this emergent phenomenology 
in the field of multi-agent complex systems.

\end{abstract}


\end{frontmatter}

\section{Introduction}

One of the main objectives of Traffic Analysis is to identify predictive models. Models are so important because
they make available practical tools (such as traffic simulators) that are widely required by TelCos
(Telecommunication Companies) in core areas such as Network Engineering and Marketing. In addition, those models
can supply mathematical and physical insights into the phenomenon's nature, giving a better understanding,
providing generalist views and thus, making possible valuable innovation.

Up to today, a lot of efforts has been made to typify different aspects and qualities of the Internet traffic:
flows ~\cite{paxon1,wilder,brownlee}, TCP connection arrivals ~\cite{paxon2}, size of FTP files ~\cite{downey},
Web transfers ~\cite{pitkow}, etc. However, there are few works focused on modeling the user itself
~\cite{crovella,vicari}. In fact, there is not much perspective of the Internet as a collection of multiple
human-driven agents (partially due to its size and ever changing nature). As a consequence, this could be an
interesting perspective, specially if it could shed some clues of globally efficient, emergent or self-organized
behavior.

From this point of view, this paper describes the process of Traffic Analysis and Modeling of an Internet Access
Link taking the user as a reference. We consider the campus Network of the Public University of Navarre (UPNA).
It represents a Local Network connected to the Internet with a typical and heterogeneous traffic profile and a
community of several thousands of users. We explore the use of the Internet Access Link and infer a new
statistical user model based on the probabilistic traffic demand observed for individual users. In order to
identify the characteristics and origins of this particular model, an statistical data analysis is further
performed.

The paper is structured as follows. Section 2 describes the measurement environment and the traffic traces
collected for this work. Section 3 discusses the global behavior observed in the Network and reveals the
co-existence of different patterns of behavior in the users' community. Section 4 presents the mathematical
modeling applied to each of these patterns (or types of users). And as a conclusion, Sections 5 and 6 obtain the
global model and discusses its significance.

\section{Measurement Environment}

We consider the campus Network of the Public University of Navarre
(UPNA) in Pamplona (Spain), which is connected to the Internet by
two Access Links coming out from a border router. At the time of
the measurements (Dec.-2003 and Apr.-2004) those were two ATM OC12
links with a normal use of 10\% of the Bandwidth and peaks of
25\%. There were a minimal set of filters active in the firewall
rules for specific applications: specially for applications
designed for use in the local network, some security limitations
on the Mail (SMTP), and an incipient filtering on individual ports
trying (unsuccessfully) to prevent Peer-to-Peer traffic).
Therefore, a mostly unrestricted access to the Internet was
granted for every user.

A passive network monitor, placed in the Fast Ethernet Link to the border router, was used to sniff and record
 the traffic packets on normal operation of the Network. Software-based measurement tools were used to passively copy
the traffic and build a trace file. The measurement enviroment is described in Fig.~1 with the details of
downstream and upstream directions for the captured traffic.

\begin{figure}[h]
\centerline{\includegraphics[width=9cm]{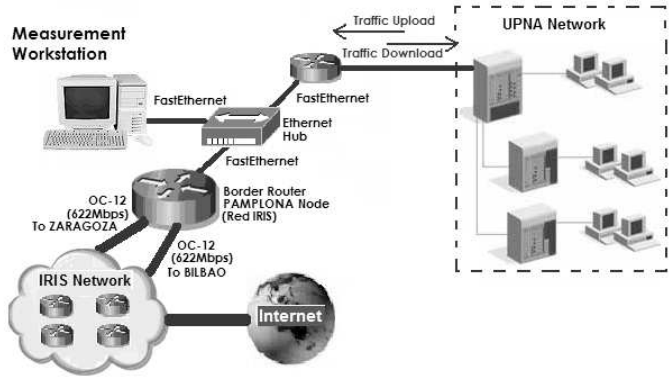}} 
\caption{Details of the Measurement Scenario.} \label{fig1}
\end{figure}

It consists of a fast-ethernet hub and a commodity workstation with GNU/Linux Operating System and the Packet
Capture Library (Lib-pcap,~\cite{pcap}) installed, to provide packet capture capability. A validation is
performed on these tools to ensure that no packet losses occur during the process. The post-process data
analysis is done by means of ad-hoc developed applications. The details of the collected packet traces can be
seen in Table~1.

\begin{table}[h]
\begin{center}
\begin{tabular}{|c|c|c|c|c|c|c|c|}
\hline  $Day$ & $Date$ & $MDtg (Down)$ & $GBytes (Down)$ & $MDtg (Up)$ & $GBytes (Up)$\\
\hline $1$ & $We\; 2003/17/12$ & 71.06 & 41.86 & 62.58 & 18.80 \\
\hline $2$ & $Th\; 2003/18/12$ & 79.21 & 44.79 & 67.69 & 19.32 \\
\hline $3$ & $Mo\; 2004/19/04$ & 68.75 & 43.31 & 68.02 & 20.55 \\
\hline $4$ & $Tu\; 2004/20/04$ & 74.31 & 42.69 & 89.33 & 22.40 \\
\hline $5$ & $We\; 2004/21/04$ & 71.73 & 41.05 & 75.65 & 21.26 \\
\hline $-$ & $Total\;Traffic$ & 365.07 & 213.73 & 363.28 & 102.35 \\
\hline
\end{tabular}
\end{center}
\caption{Detail of traffic traces used in the study. The Number of
IP Datagrams (in Millions, MDtg) and the amount of Bytes (in
Gigabytes, GBytes) are given for each day of measurement.}
\end{table}

They comprise 5 complete (24 hours/each) normal working days, belonging to two different periods of the year
(Dec.-2003 and Apr.-2004) in order to also measure the time evolution of the model. These traces involved 728.36
Mill. of IP Datagrams and 316.08 Gigabytes of Total Traffic. The data bytes in the IP Datagrams include the
counting of the header bytes.

The monitored traffic presents a standard time profile and
composition, with typical daytime patterns of activity, TCP
protocol predominance (98\% of total traffic Bytes) and asymmetric
traffic (70\% Downstream, 30\% Upstream). As a consequence, the
community under study shows some characteristics that are typical
of an ordinary community of users ~\cite{paxson3}, a medium-sized
Intranet connected to the Internet in flat rate mode.

\section{Traffic Analysis in the Community of Users}
\subsection{The Global View}
First, the global behavior of the entire population is analyzed.
As it was shown in Section 2, TCP comprises the most significant
load of traffic registered. Moreover, total traffic is dominated
and shaped by the download component. These effects will lead us
to focus on TCP downstream traffic for modeling.

If the whole community of users is ordered according to their
daily traffic activity, we obtain an enlightening global picture
(see Fig.~2).

\begin{figure}[h]
\centerline{\includegraphics[width=6cm]{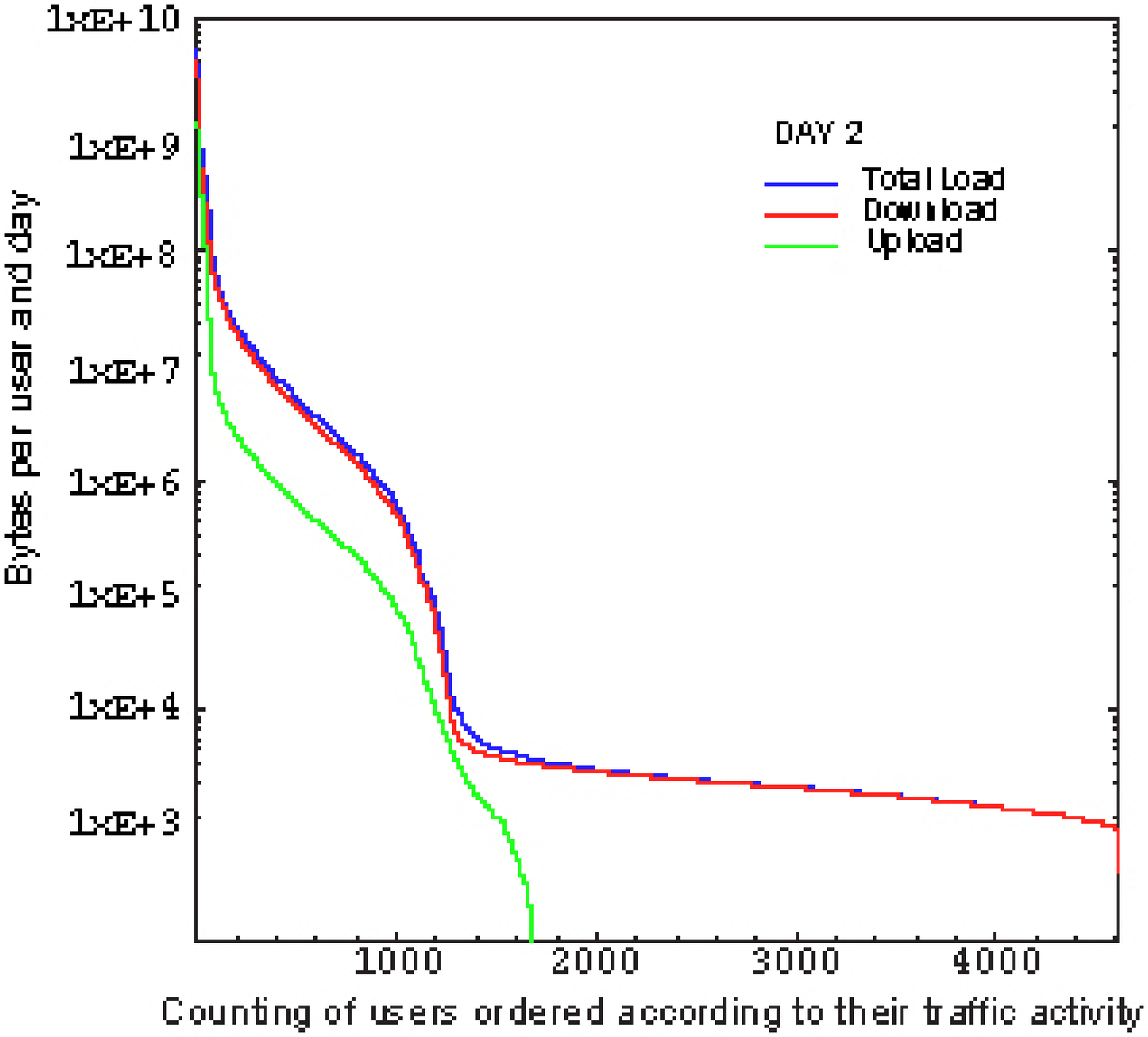}\hskip
10mm\includegraphics[width=6cm]{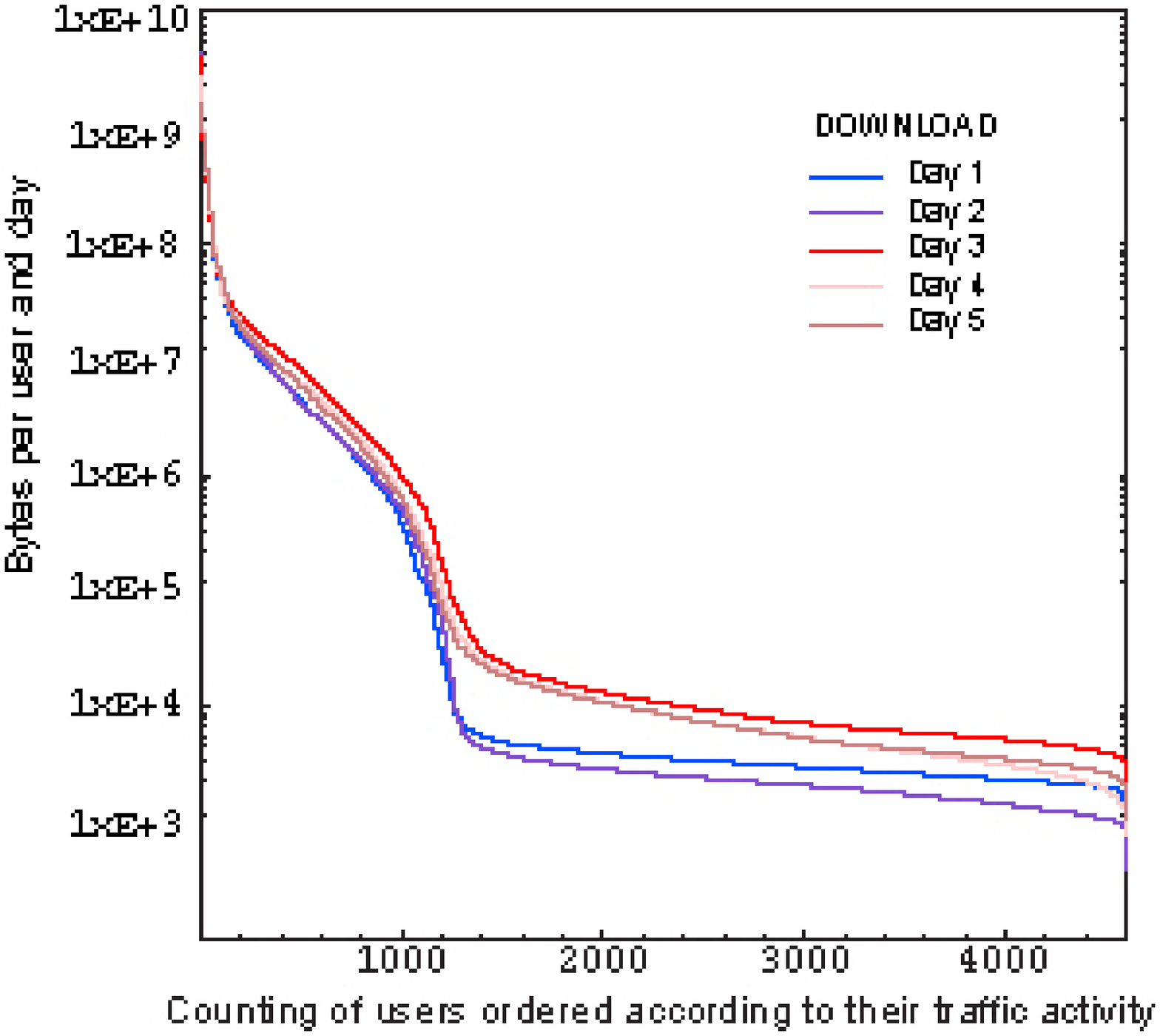}} \centerline{(a)\hskip
6cm (b)} \caption{(a) Bytes of TCP traffic registered per user on
Thursday 2003/18/12 (total, down and upstream components)(b)
Downstream Bytes of TCP traffic registered per user for all days.}
\label{fig2}
\end{figure}

Fig.~2a shows the TCP traffic registered per user on Day 2 with
its downstream and upstream components. Here, the dominance of the
downstream component over the total traffic amount is observed,
the same way as the upstream is found to have a minor weight.
Users are ordered in the horizontal axis according to their
traffic loads (total, down or up respectively). Thus, the first
user on this axis is the user with the heaviest traffic activity
and user number 4608 is the user with the lowest activity in the
community. These users can be different persons for each
component, as the most intensive user in the Download is typically
a different person than the most intensive user in the Upload
component.

Fig.~2b shows the TCP traffic downloaded per user and day for
all traces (2 days in Dec.-2003, depicted in blue and 3 days
Apr.-2004, shown in red). A positive drift is observed in the
consumption among the traces spaced in four months, but the shape
of the plots remains constant. Total and Upstream traffic are
removed from this figure for clarity but they also show similar
shapes in all traces.

Two main characteristics are observed in Fig.~2:

1.- Two inflection points and three different regions become visible in the graphics (Fig.~2b). The first point
is found at the value of around 30 MB/day for all traces, and the other one at the values of 10 or 30 KB/day for
Dec.-2003 or Apr.-2004, respectively. The three areas are found in all the traces. A longitudinal variation
gives information of a tendency of increasing consumption for the lighter users as months pass, but the shapes
of the curves remain constant for the whole community. Each inflection point represents a mark at which the
behavior of users' consumption changes dramatically. These findings reveal that somehow all users don't behave
uniformly. In fact, they can be classified in three clusters of different statistical behavior, as we show in
the next sections.

2.- A great number of users don't generate traffic (0 Bytes/user/day in the upstream). In Fig.~2a, it can be
seen that users beyond number 1688 do not show any uploaded traffic. Their IP addresses register download
traffic because they were automatically traced from the Internet by different applications (like spiders, etc.)
but they never responded. This leads us to consider that they are not real users, but just unallocated IP
addresses or unconnected machines in the Network. These IP addresses are eliminated from our study.

\subsection{Searching for Patterns of Behavior}

As a second step, we analyze users' behavior during each day of the five collected traces. The main statistical
magnitudes of downloaded TCP traffic for users within of the three regions of consumption are obtained. These
different groups of users have invariant patterns for each one of the analyzed days. We refer to these groups as
Type A, Type B and Type C users, for the regions of the heaviest, the medium and the lightest consumers,
respectively. The data displayed in Tables~2 and 3 show the invariance of the behavior in all traffic
traces in terms of number of users, downloaded traffic per day and behavior during the day.

\begin{table}[h]
\begin{center}
\begin{tabular}{|c|c|c|c|c|c|c|}
\hline $Object$ & $Statistical\;Magnitude$ & $Day\; 1$ & $Day\; 2$ & $Day\; 3$ & $Day\; 4$ & $Day\; 5$\\
\hline $Total\; Users$ & $Number\; of\; Users$ & 1727 & 1688 & 1782 & 1766 & 1775\\
\hline $Total\; Traffic$ & $Downloaded\; GBytes$ & 40.60 & 41.37 & 42.40 & 41.63 & 40.35\\
\hline      & $\%\; Users$ & 7,01\% & 8,12\% & 8,36\% & 7,42\% & 7,94\%\\
\cline{2-7} $TYPE\; A$ &  $\%\; Download$ & 87,07\% & 87,48\% & 82,60\% & 84,41\% & 85,44\%\\
\cline{2-7} & $<X>$ & $89MB$ & $78MB$ & $63MB$ & $91MB$ & $75MB$\\
\cline{2-7} & $\sigma$ & $541MB$ & $527MB$ & $472MB$ & $422MB$ & $366MB$\\
\hline      & $\%\; Users$ & 65,32\% & 66,59\% & 67,62\% & 66,25\% & 63,83\%\\
\cline{2-7} $TYPE\; B$ & $\%\; Download$ & 12,92\% & 12,52\% & 17,39\% & 15,57\% & 14,54\%\\
\cline{2-7} & $<X>$ & $2.08MB$ & $2.01MB$ & $3.04MB$ & $2.72MB$ & $2.55MB$\\
\cline{2-7} & $\sigma$ & $6.57MB$ & $6.69MB$ & $7.69MB$ & $7.30MB$ & $6.75MB$\\
\hline      & $\%\; Users$ & 27,68\% & 25,30\% & 24,02\% & 26,33\% & 28,23\%\\
\cline{2-7} $TYPE\; C$ & $\%\; Download$ & 0,01\% & 0,004\% & 0,01\% & 0,01\% & 0,01\%\\
\cline{2-7} & $<X>$ & $4.47KB$ & $3.11KB$ & $10.69KB$ & $7.77KB$ & $7.79KB$\\
\cline{2-7} & $\sigma$ & $1.53KB$ & $1.65KB$ & $6.55KB$ & $7.03KB$ & $6.77KB$\\
\hline
\end{tabular}
\end{center}
\caption{Main statistical magnitudes obtained for each region show stability over time. These magnitudes are:
the distribution of number of users and traffic consumption in each region (expressed as a percentage of the
total magnitudes for each day) and the levels of consumption per user and day, expressed as the mean value $<X>$
and the standard deviation $\sigma$ of the downloaded traffic (Bytes) for the cluster of users in every region.}
\end{table}

In Table~2 we show that the number of users in each of the three regions of Fig.~2 and its global
consumption per day are almost invariant for all traces. These magnitudes are expressed as a percentage over the
total number of users and total bytes of traffic per day, in order to make it clearer. This is also true for the
levels of mean consumption per user and day and the standard deviation of the statistical sample. So, each type
of user represents a group of individuals that is almost constant in number and traffic demanded per day.

Table~3 describes the pattern of consumption for each group during day 1 (Wednesday 2003/17/12). This is
another invariant property observed in the registered traces and it shows a characteristic daytime pattern for
each group. In region A, users have a constant and heavy demand over day and night. Users in region B show a
stronger consumption in the mornings. In region C the traffic registered is very low and similar during the
whole day.

\begin{table}[h]
\begin{center}
\begin{tabular}{|c|c|c|c|c|}
\hline $Type\; of$ & $Morning$ & $Afternoon$ & $Total\; Day$ & $Night$\\
       $User$ & $(9h/13h)$ & $(13h/18h)$ & $(9h/18h)$ & $(18h/9h)$\\
\hline $TYPE\; A$ & 26,63\% & 27,82\% & 54,45\% & 45,55\% \\
\hline $TYPE\; B$ & 41,16\% & 32,75\% & 73,91\% & 26,09\% \\
\hline $TYPE\; C$ & 28,25\% & 23,14\% & 51,39\% & 48,61\% \\
\hline
\end{tabular}
\end{center}

\caption{Distribution of consumption during Day 1 (Wednesday
2003/17/12) for the three observed groups. The daily distribution
of downloaded traffic can be observed in its day-night
composition. Also the day components are disaggregated with the
morning-afternoon detail.}
\end{table}

\subsection{Three Clusters of Users}

The analysis sketched in the previous subsection gives us the portrayal of a community with three different
types of users (A, B and C) coexisting in the Internet Access. Clustering them in three different groups allows
us to highlight the size disparity in traffic consumption. Size disparity is often called the ``{\it mice}-{\it
elephants}'' phenomenon, traditionally observed in Internet flows (according to duration or bit-rate), but also
referred as volume of traffic \cite{broido1}. An object is labeled as {\it mouse} if it contributes to the
number of total objects of an ensemble, {\it elephant}, if it contributes to the total volume mass and in
~\cite{broido2} hosts in between are called {\it mules}. Finally we name each type of users after this
terminology. The main features of each group are:
\begin{enumerate}
    \item Type A users (or ¨{\it ELEPHANTS}): They are the great consumers, with a heavy and
    timely constant demand of Internet traffic. They take the 90\% of the total traffic (see Table~2),
    and they have a continuous activity during the day, with almost 50\% traffic at night and 50\% at day (see Table~3).
    \item TYPE B users (or {\it MULES}): They are the pragmatic users. Their activity on the net matches
    with business hours of activity.  It takes place more intensively in the mornings (41\%, see Table~3),
    and the total amount of day load is almost the 75\% of their total download. It seems that they use the Internet Access as a working tool.
    \item TYPE C users (or {\it MICE}): They are the erratic and vanishing users with an extremely low and
    sporadic demand of traffic, and random daily patterns difficult to be tracked.
    The total amount of traffic demanded by these users is residual (see Table~2).
    The consumption is very small and it has the same residual weight at day and at night,
    almost 50\% traffic at night and 50\% at day (see Table~3).
\end{enumerate}

This community seems to follow the Pareto Principle or the ``80-20 rule", which says that, when something is
shared among a sufficiently large set of participants, there will always be a number k, such that k\% of the
resources is taken by (100 - k)\% of the participants. In our case, k=90, meaning that the 10\% of the
population (the {\it ELEFANTS} group) is responsible for the 90\% of the downloaded traffic. For the rest of the
users, medium consumers ({\it MULES}) are the 65\% and they make a businesslike use of the resources, taking
almost the rest of the bandwidth `pie'. Low consumers ({\it MICE}) contribute only in numbers, the 25\% of the
total, and they have residual and apparently random traffic demands.

Let us remark at this point that many complex systems present similar generic features that include the
Pareto-like statistical behavior. The most popular examples can be found in areas such as econophysics (e.g.,
distributions of wealth or income in western economies or standardized price returns on individual stocks
markets, Company size distribution ~\cite{dragulescu,leong,ramsdena}), social networks (e.g., size distribution
of human settlements, frequency of human relations ~\cite{roehner,liljeros}), scale-free networks (Epidemic
spreading and transportation dynamics ~\cite{satorras,guimera}), natural phenomena (e.g., Gutenberg-Richter law
of earthquake magnitudes ~\cite{guttenberg}), or Biolgy (protein sequence length distribution ~\cite{jaina}).
Hence, some similarities appear between all these systems and the distribution of downloaded Internet traffic
for users of an Intranet.

\section{Models for the Users' Downloaded Traffic}

As explained before, the particular users conducting alike are grouped in three different clusters and a
mathematical model is built here for each region of Fig.~2b. Thus, the global model will be obtained as the
mixture of three different users' models.

If the real-valued random variable $X$ represents the amount of incoming TCP traffic that a user consumes per
day, expressed in bytes, the downstream traffic belonging to each group is used to build three different
Cumulative Distribution Functions (CDFs). In our case, X attains discrete values $x_i$, with probability $p_i =
p(x_i)=P(X=x_i)$. The values of $p_i$ are calculated as the number $n_i$ of users that consumed $x_i$ bytes
divided by the number $N_c$ of total of users in the same cluster.
\begin{equation}
F(x) = P(X\leq x)=\sum_{x_i\le x} P(X=x_i)= \sum_{x_i\le x}
p(x_i)= \sum_{x_i\le x}{n_i\over N_c}.
\end{equation}

As we are interested in the probability that a user downloads more than x bytes/day, we will work with the
Complementary Cumulative Distribution Function (CCDF), which is defined as:
\begin{equation}
F_c(x) = P(X>x)=1-F(x).
\end{equation}

In the next subsections, we select three CCDFs to model the downstream TCP traffic for each different group and
we fit our monitored real data to these CCDFs. We will begin with TYPE A users, continue with TYPE C and end up
with TYPE B users, for they are going to build the bridge between the two extremes.

\subsection{Model for Type A Users}

This cluster represents around the 10\% of total users, consuming almost 90\% of the traffic (see Table~2).
It follows the Pareto Principle. If we move in the x-axis towards user number one in Fig.~2b, it can be seen a
staggering increase in the traffic demand per user. So a fast variation is observed only for this cluster and
intuitively could give some clues of a long-tail behavior in the CCDF. If we plot the measured CCDF data in a
double logarithmic plot we observe a straight line arrangement for the data-points, which is consistent with a
power-law dependence. Based on these observations, a Pareto distribution is used to model this group.

\begin{equation}
F_c(x) = b*x^{-\alpha},   \hskip 1cm    b\ge0,\hskip 2mm \alpha\ge0, \hskip 2mm x\ge b^{\frac{1}{\alpha}}.
\end{equation}

A double logarithmic plot for the 5 days is done in Fig.~4 to show the details of the fitting. The data follow
a two straight line arrangement and the approach of fitting the empirical data to two Pareto distributions is
selected, as it seems to be more general and complete than selecting a truncated Pareto as CCDF (see ~\cite{guimera}).

The first line of data is labeled as ``Pareto 1'' in Fig.~3a and 3b and appears in the area of transition
between TYPE A and TYPE B users. This transition exists because the threshold value that separates regions A and
B (30MB/user/day) is a  point, that is chosen among an area of inflexion points.

\begin{figure}[]
\centerline{\includegraphics[width=12cm]{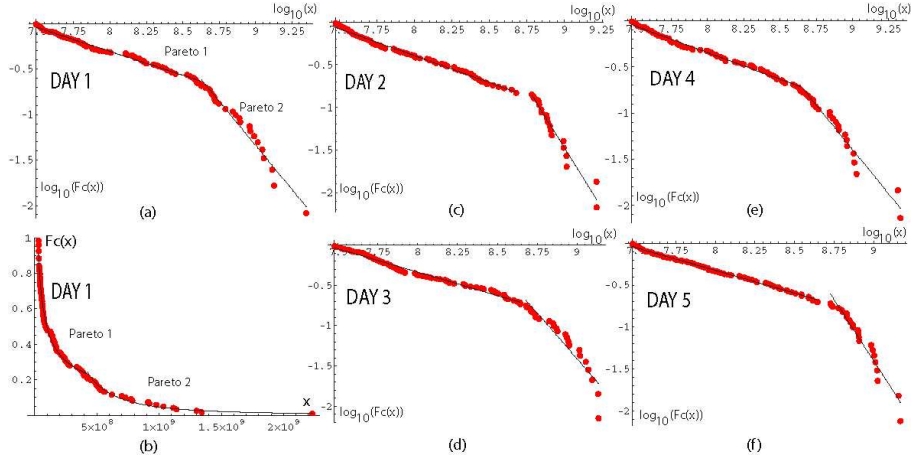}}
\caption{(a,c,d,e,f) Graphic details of the fittings for
ELEPHANT users in double logarithmic plot for all the traces.
(b) Empirical and fitted CCDF  obtained for day 1.}
\label{fig4}
\end{figure}

The result of finding two straight lines and no just one, indicates that there is not a well-defined threshold
or boundary that separates TYPE A users from the rest (observed in ~\cite{liljeros}). It also indicates that
there are two speeds of consumption within the {\it ELEPHANTS} herd, as the power law behavior has different
characteristics in each distribution. The ``Pareto 1'' zone has an exponent $\alpha$ $<1$ and the value of
$\alpha$ for the ``Pareto 2'' is around 2. Table~4 shows the details of the fitting parameters.

\begin{table}[h]
\begin{center}
\begin{tabular}{|c|c|c|c|c|c|}
\hline $DAY$ &$1$ & $2$ &$3$ & $4$ & $5$\\
\hline $b_1$ &$7143.08$ & $44700.9$ &$19500.0$ & $18613.8$ & $10242.7$ \\
\hline $\alpha_1$ &$0.519$ & $0.630$ &$0.578$ & $0.576$ & $0.541$ \\
\hline $b_2$ &$1.45E+16$ & $1.25E+24$ &$1.09E+19$ & $9.25E+15$ & $8.06E+24$ \\
\hline $\alpha_2$ &$1.944$ & $2.842$ &$2.272$ & $1.928$ & $2.92$ \\
\hline
\end{tabular}
\end{center}
\caption{Values of the fitting parameters b and $\alpha$ obtained for the five days
in the two zones ``Pareto1'' and ``Pareto2''.}
\end{table}

 These users consume a heavy load of traffic every day. Consumption varies in a wide range between users (10 GBytes and 30 MBytes per user and day). In Table~2 the mean value of incoming traffic is
around 80 MBytes/user/day but with a high dispersion value, typical of long-tail distributions (the standard
deviation for this population is around 400-500 MBytes, revealing a high spread of its values). They are always
on-line and perform an intensive use of the link, with a constant and uniform traffic profile over day and night
(see Table~3). They download high quantities of data, probably heavy Web- or FTP-files, or even they can be
connected to P2P networks.

\subsection{Model for Type C Users}

Users belonging to this group consume a residual quantity of traffic, with random and irregular patterns of
activity (see Table~2 and 3). They show a better marked mean consumption than {\it ELEPHANTS}, and plotting
the measured CCDF data in a natural logarithmic plot a straight line arrangement appears for the data-points,
which is consistent with an exponential dependence. Finally the exponential distribution is used to model this
group:

\begin{equation}
F_c(x) = b e^{-ax},   \hskip 2cm    a\ge0,\hskip 2mm x\ge0.
\end{equation}

Fig.~4 shows the details of the fitting in a plot that represents y-axis in natural logarithm for the 5 days.
Table~5 shows the details of the fitting parameters.

\begin{figure}[h]
\centerline{\includegraphics[width=12cm]{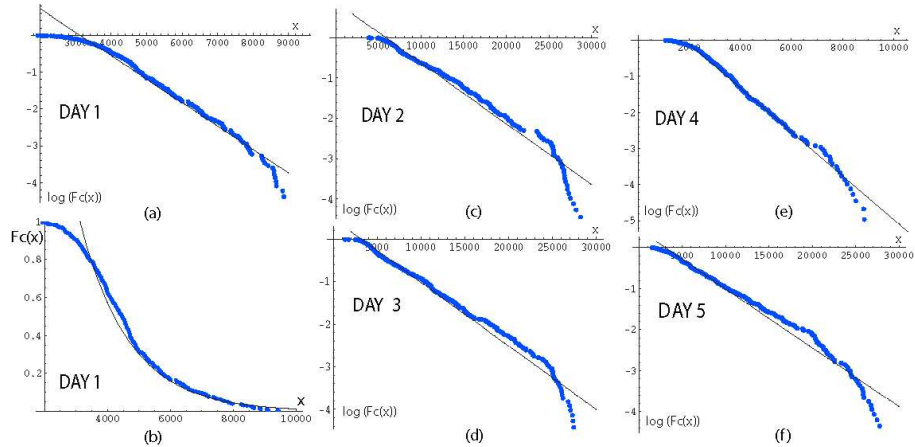}}
\caption{(a,c,d,e,f) Graphic details of the fittings
for {\it MICE} users in natural logarithm plot for all days.
(b) Empirical and fitted CCDF  obtained for day 1.}
\label{fig44}
\end{figure}

\begin{table}
\begin{center}
\begin{tabular}{|c|c|c|c|c|c|}
\hline $DAY$ &$1$ & $2$ &$3$ & $4$ & $5$\\
\hline $a$ &$6.37E-4$ & $1.50E-4$ &$1.49E-4$ & $6.05E-4$ & $1.43E-4$ \\
\hline $b$ &$7.32$ & $2.41$ &$1.61$ & $3.01$ & $1.53$ \\
\hline
\end{tabular}
\end{center}

\caption{Values of the fitting parameters b and a obtained for the five days}
\end{table}

The volume of consumption of {\it MICE} is so low that one could question the worth of modeling them. But let us
recall that it is an important population within the Network, around the 25\% of the users belong to this group.
They contribute to the number of entities but not to the volume. The demand of traffic is below tens of
KBytes/user/day, with a weak tendency of growth as time passes, from threshold values of 10KB in Dec.-2003 to
30KB in Apr.-2004 (see Fig.~2b). This means that just a hundred of packets (40-1500 bytes long) have been
received per day by one of this users, typical for sporadic surfing or mail applications activities.

\subsection{Model for Type B Users}

Most users belong to this group (approximately 65\%, see Table~2). Their consumption lies somewhere in
between the last two extremes. The Weibull distribution can offer a way of transition from an exponential to a
heavier-tailed distribution with a shape parameter ($\beta<1$), and it is found in literature that some
empirical distributions are often close to this type of distribution \cite{broido1}. Hence, a Weibull
distribution is selected in this case as CCDF, due to its flexibility and because it can mimic a transition from
the exponential to the Pareto distribution.
\begin{equation}
F_c(x) = e^{-ax^{\beta}}, \hskip 2cm    a\ge0,\hskip 2mm \beta\ge0,\hskip 2mm x \ge 0.
\end{equation}

The details of this approximation are shown in Fig.~5 in a logln-log plot that represents the decimal and
natural logarithm of $F_c(x)$ in the y-axis and the decimal logarithm of $x$ in the x-axis for the 5 days of
data collection. Table~6 shows the details of the fitting parameters.
\begin{figure}[h]
\centerline{\includegraphics[width=12cm]{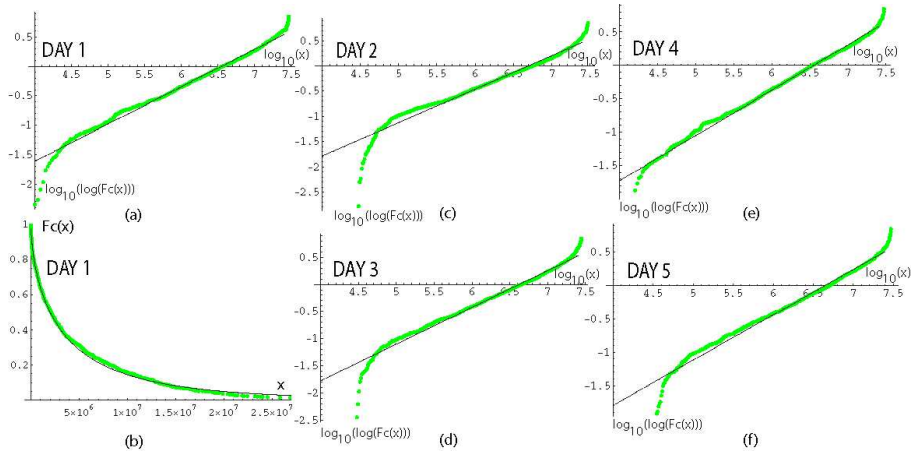}}
\caption{(a,c,d,e,f) Graphic details of the fittings for {\it MULE}
users in logln-log plot for all days.
(b) Empirical and fitted CCDF  obtained for day 1.}
\label{fig5}
\end{figure}

\begin{table}[h]
\begin{center}
\begin{tabular}{|c|c|c|c|c|c|}
\hline $DAY$ &$1$ & $2$ &$3$ & $4$ & $5$\\
\hline $a$ &$7.08E-5$ & $3.59E-5$ &$3.33E-5$ & $3.62E-5$ & $3.17E-5$ \\
\hline $\beta$ &$0.633$ & $0.663$ &$0.675$ & $0.678$ & $0.676$ \\
\hline
\end{tabular}
\end{center}

\caption{Values of the fitting parameters b and a obtained for the five days.}
\end{table}

This group of users belong to an intermediate kind (see Table~2), with a moderate consumption of traffic and
use of the Internet access. A {\it MULE} digests a traffic load in the range of tens of KBytes to tens of MBytes
per user and day, showing a growing tendency for the lighter consumers as time passes. The average demand per
user is around 2 or 3 MBytes/user/day (in Dec.-2003 and 7 in Apr.-2004, respectively) and the dispersion value
is high (the value of the standard deviation is around 6-7 MBytes/user/day), getting closer to the {\it
ELEPHANT}'s profile. It is also observed in Table~3 that they concentrate their activity in business hours,
possibly using the Internet Access in their work activities. These users seem to read emails, make sporadic
surfing on the net and download relatively light Web pages or files (email attached or FTP).

\section{The General Model}

We proceed now to propose a statistical user model to describe the incoming traffic in
the Internet Access for the general community. This is the result of the combination of the
three models which were obtained for each type of user.

The CDF function F(x) for this global model is plotted in Fig.~6, showing the empirical data registered on Day
1 (Wed. 2003-17-12) and the theoretical distributions fitted to these data. This global CDF represents the
probability that a user demands less than an specific traffic load. This probability is calculated from the CDF
for the type of user (or region of consumption) averaged by the relative weight of this type of user in the
community (averaged by the number of users of each type shown in Table~2).

\begin{figure}[]
\centerline{\includegraphics[width=8.5cm]{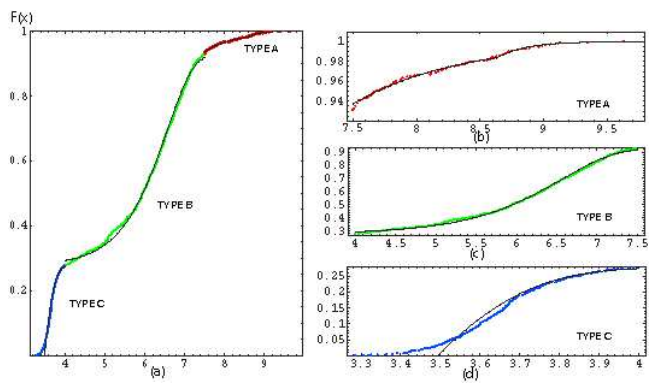}} \caption{(a) The general model built from the
combination of the individual models for each type of user (empirical data and model curves). The x-axis is in
logarithmic scale. (b) Details of the {\it ELEPHANTS} region. The two Pareto CDF approach used in the fitting
procedure can be observed here. (c) Details of the {\it MULES} region. (d) Details of the {\it MICE} region.}
\label{fig6}
\end{figure}

The three different groups of users are very well identified in the general model. We can think of the Network
as a multi-agent system in an active environment. At first sight, the agents could seem to be in a weak
competition for the bandwidth in the Internet Access, but it is more realistic to think that they are
non-interacting agents without any mutual information among them. Even under these circumstances, and
surprisingly, the system self-organizes with the pattern of behavior given in Fig.~6.

The part of the population that receives the highest traffic income, generally less than 10\% of the
individuals, follows a power-law distribution (Pareto behavior).  This means that the consumption is dominated
by a few individuals and, for the misfortune of planning engineers, that there always exists a significant
probability or risk of finding a more intensive user in the Network. In the range of low traffic download, the
cumulative distribution follows the Boltzmann-Gibbs law (exponential behavior with $F_c(x)\propto\exp(-x/T)$).
Hence, by analogy with statistical physics, one may expect that the activity of these users is characterized by
some kind of random access to the Net (memoryless), where the effective traffic `temperature' (with T= 1/a in
Table~5) would be related to the average amount of bytes consumed per agent.

Let us remark that these two regimes, the Pareto and the Boltzmann-Gibbs statistical patterns, can be found in
other real systems which also depend on the human behavior, such as the distribution of wealth or income in the
western societies ~\cite{dragulescu}. But also a third different regime is observed here, namely the group of
medium consumers or {\it MULES} described by the Weibull distribution function. This distribution can model the
transition from an exponential to a heavier-tailed distribution by varying its shape parameter ($\beta<1$). The
low traffic demand dominates this group, but the consumption increases rapidly for the last users that are close
to the Pareto-like region.

Summarizing, the model depicted in Fig.~6 collects and brings out all the information needed for the
statistical forecasting of consumption that a community of users in a Intranet will perform of the Internet
Access Link. As it is, it could also be useful for TelCos, to investigate new strategies in Network Planning or
Marketing areas of Internet consumption.

\section{Conclusions}

This work describes the process of Traffic Analysis and Modeling of an Internet Access Link taking the user as a
reference. A new probabilistic global model is obtained to describe user traffic consumption in terms of bytes
per user and day.

Considering the Network as a multi-agent system, an emergent statistical structure is found for the Internet
demand within the community. The identification of three different groups of users and the probabilistic
distributions that describe their behavior have been established.

Some particular properties are present in these experimental results. Apart from the usual Pareto and
Boltzmann-Gibbs behaviors, which are also found in other multi-agent systems commanded by the human action, a
new emergent regime is found here. This can be modeled by the Weibull distribution, that could be interpreted as
a bridge between the Pareto and the Boltzmann-Gibbs regimes.

Let us stand out that different models and real systems with dynamical processes and local interactions taking
place at the microscopic scale show similar macroscopic characteristics to those observed in this work. This
similar pattern of behavior,that continues to emerge in  our analysis, is somehow striking since it would be
more realistic to think that the agents are independent and non-interacting when they access to the Internet
link.

We hope that the findings presented in this paper will help for a better understanding of the laws governing the
interactions in multi-agent systems, and that they will provide empirical views to achieve enhanced modeling for
complex systems.

{\bf Acknowledgements:} C. P.-L. and D. M. thank UPNA (Public University of Navarre) for allowing them to
monitor the data in its Internet Access Link.


\end{document}